# Analysing ocular parameters for web browsing and graph visualization

Somnath Arjun[1], Kamal Preet Saluja[1], Pradipta Biswas[1]

[1] Centre for Product Design and Manufacturing,
Indian Institute of Science,
Bangalore
`somnatharjun@iisc.ac.in`
`kamalpreets@iisc.ac.in`
`pradipta@iisc.ac.in`

**Abstract.** This paper proposes a set of techniques to investigate eye gaze and fixation patterns while users interact with electronic user interfaces. In particular, two case studies are presented - one on analysing eye gaze while interacting with deceptive materials in web pages and another on analysing graphs in standard computer monitor and virtual reality displays. We analysed spatial and temporal distributions of eye gaze fixations and sequence of eye gaze movements. We used this information to propose new design guidelines to avoid deceptive materials in web and user-friendly representation of data in 2D graphs. In 2D graph study we identified that area graph has lowest number of clusters for user's gaze fixations and lowest average response time. The results of 2D graph study were implemented in virtual and mixed reality environment. Along with this, it was observed that the duration while interacting with deceptive materials in web pages is independent of the number of fixations. Furthermore, web-based data visualization tool for analysing eye tracking data from single and multiple users was developed.

## 1. Introduction

Information visualization is one of the major approaches to analyse data. A great deal of research on information visualization have been implemented in last two decades. With ever-increasing amount of available computing resources and sensing devices, ability to collect and generate a wide variety of large, complex datasets continues to grow. As a result, visualizing and analysing those complex datasets becomes challenging. Although an ample amount of visualization techniques has been devised for a wide variety of applications, they tend to follow one-size-fits-all paradigm with respect to users. The designs are rarely personalized to a specific user characteristic or type of dataset. Data adaptive visualization and personalization of user interfaces are our long-term research goal. Considering our research goal, eye gaze and fixation patterns while users interact with electronic user interfaces was investigated. Eye tracker is a device which can capture eye gaze movements and viewer's gaze on a stimulus. There are different kinds of eye movements like fixations, saccades, pupil dilation and scan paths [1]. The eye movements provide unique insight into visual search task and can be analysed to derive user's attention patterns. We presented two case studies

### 1.1 Analysing graph visualization

One of the fundamental and orthodox way of visualizing data are with graphs and charts. Till today static graphs and charts are used for visualizing data in substantial amount. Appropriate graph for dataset is difficult to choose as it depends on features of the data and problem statement. With data visualization, issues like huge volumes of data and limitations of algorithms can be confronted easily. Visualization techniques and tools have been developed in abundance to mitigate the above-mentioned scenarios. Despite these tools and techniques, optimizing visualization and interaction techniques still poses research challenges. Static graphs in 2D were considered for the study and results of this study was implemented in 2D and virtual reality environment which is described in section 4.

### 1.2 Analysing deceptive materials in web pages

It is in practice to design deceptive user interfaces to manipulate users by exploiting human psychology. Web pages have these deceptive materials which tricks user to do tasks that they did not intend to. There are no easy solutions or alternatives to these deceptive patterns. Nir Eyal, in his book [2] explains how a good understanding of cognitive science can add value to user understanding. Despite these possible solutions, there are still space for significant improvement. With the aim of providing solution to this issue, a study was undertaken for analysing eye gaze and fixation patterns while user interact with web pages having deceptive materials.

The above two case studies exhibit two different aspect of helping designer through eye gaze data analysis but still getting access to an eye gaze tracker is not always possible. For example, cheaper eye gaze tracker like Tobii Eye-X model have restrictions in terms of research purpose, hence we propose webcam-based analysis tool in section 5. The tool is independent of the implementation of webcam-based eye gaze tracker and can generate similar visualization graph as reported in the study.

## 2. Related Work

Information visualization research has maintained one-size-fits-all approach, ignoring an individual user's need, types of data and domain of applications. Ziemkiewicz et.al. [3], as well as Green and Fisher [4] have shown that the personality trait of locus of control can impact relative performance for visualizations. These results indicate that there is an opportunity to apply adaptation and personalization to improve usability. One of the attempts to adapt to individual user differences in visualization is presented in [5]. A user's visualization expertise and preferences are dynamically inferred through monitoring visualization selection (e.g. how long it takes a user to decide on which visualization to choose). Steichen et. al. [6] analysed sequential nature of user eye gaze patterns and found several gaze behaviours differences between different user/task groups during information visualization usage. These results could be leveraged by adaptive information visualization systems in order to automatically identify different user and task characteristics. In this paper similar kind of study was conducted as Stienchen's [6] for system generated data with the goal of developing optimized visualization techniques for smart manufacturing in both 2D and 3D. Existing visualization techniques for smart manufacturing explored representing relationship among data through establishing ontologies and visualizing network diagram among different items [7]. Sackett [8] presented a review on existing visualization techniques but did not provide detail on visualizing both temporal and spatial information simultaneously. Deceptive user interfaces are interfaces that tricks user to do tasks that they did not intend to. The effect these patterns have on users varies from offensive to subtle or no influence. Fogg [9] described deceptive patterns as techniques that are used to obtain unintended outcome. Gray's taxonomy [10] of deceptive materials is based on the strategic motivator behind patterns. Interface interference that gray suggested broadly refers to visual and language manipulation. Visual manipulation is a technique where the image, components in the image or visual cue is maneuvered to puzzle user for completing their intended task. Language manipulation refers to writing confusing statement in user interface copy or guilt tripping the user. Rosis [11] stated that these deceptive patterns are both unintentional and on purpose from designers of user interfaces. An eye tracking study was conducted earlier to measure how including text and pictures affects online reading [12]. Pan et. al. [13] undertook study to explore the determinants of ocular behaviour on web pages. In this paper, a similar study was undertaken to explore severity of impact on users while executing task on web pages with deceptive patterns.

## 3. User Study

To investigate eye gaze and fixation patterns while users interact with electronic user interfaces, a user eye tracking study was designed for 2D graphs and deceptive materials on web pages.

### 3.1 Study on 2D graph visualization

**Aim of the study:** A study was undertaken to investigate how users interpret information from graphs. New technique for analysing gaze data was developed based on soft clustering. In particular, expectation maximization (EM) algorithm was investigated. XB cluster validation index [14] was used for validating optimum number of clusters. Using these soft clustering techniques, the number and locations of areas of interest in a visual display was automatically identified. This study will be useful to point anomalies in the current visualization techniques.

**Expectation–Maximization** (EM) is an iterative method to find maximum likelihood or maximum a posteriori (MAP) estimates of parameters in statistical models, where the model depends on unobserved latent variables. The EM iteration alternates between performing an expectation (E) step, which creates a function for the expectation of the log-likelihood evaluated using the current estimate for parameters, and a maximization (M) step, which computes parameters maximizing the expected log-likelihood found on the $E$ step. These parameter-estimates are then used to determine the distribution of the latent variables in the next E step.

**XB cluster validation index:** A cluster validity function proposed by Xie and Beni [14] is used to evaluate the fitness of partitions produced by clustering algorithms. It is defined as the ratio of the compactness measure and separation measure, i.e. lower index value indicates fitter partitions.

**Participants:** The user study was conducted with 9 participants, among them 6 males and 3 females, everyone between 20 and 35 years.

**Materials:** A Tobii Eye-X tracker was used for recording eye gaze, 29-inch display monitor with 1366×768 screen resolution and Lenovo yoga laptop with i5 processor for conducting the user study.

**Design:** A software which consisted four basic visualization techniques and five set of questions with multiple choice answers was developed. An eye gaze tracker was placed at the bottom of the screen. Participants were asked to seat at 75 cm away from the screen. They were instructed to answer a set of questions by investigating the graph. Figure 1 below shows a sample interface of the system. Bar graph, line graph, radar graph and area graph were considered for the study.

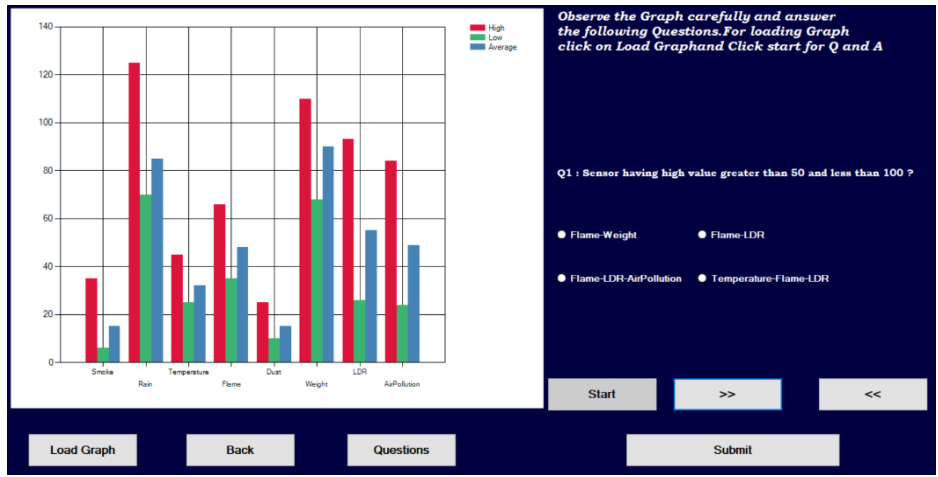

**Fig. 1.** Software interface showing graphical and question/answer portion

For each graph, the following set of questions were displayed one at a time.

**Q1:** How many sensors have lesser average value than average of all low values?
**Q2:** Average of which sensor is approximately same as the average of all sensor's average value?
**Q3:** Sensors having high value greater than 50 and less than 100
**Q4:** Two sensors reading showing nearly equal low values with minimum difference.
**Q5:** What is the approximate average of all high values of sensors?

**Procedure:** Participants were briefed about the experiment. For each participant, the eye tracker was calibrated using the 9-points calibration routine. After calibration, participants were asked to undertake the study. The x-y coordinates of the gaze location and response to each question with timestamp were logged in a text file.

**Results:** We analysed number of correct answers, average time for correct answers, total time taken for all answers and the optimal (lowest) number of clusters for user's gaze fixation. We found that bar graph had highest number of correct answers with 27 correct answers and radar graph had lowest with 19 correct answers out of 45 questions, all users cumulatively. Area graph had lowest average response time for individual questions (24.6 seconds) and total time for all questions (155.86 seconds). We undertook one-way ANOVA for all dependent variables and did not find significant differences for any of the dependent variables, [p>0.05].

### 3.2 Study on deceptive materials of web pages

**Aim of the study:** An eye tracking based study was undertaken to investigate deceptive materials on web pages. Three types of deceptive patterns were considered, visual manipulation, language manipulation and combination of visual and language manipulation.
**Participants:** 11 participants were involved in the experiment: 8 males and 3 females, with the average age of 24.
**Materials:** The study was undertaken with infrared camera-based Tobii Pro X3-120 eye-tracker [15] system with 120 Hz sampling frequency. Tobii studio software working under MS Windows10 (x64) was used along Lenovo yoga laptop with i5 processor.
**Design:** A software which consisted twelve screenshots of different web pages with deceptive techniques was developed. It also contained a set of questions with multiple choice answers. Out of the twelve images, three images were based on type-1(visual manipulation), three on type-2(language manipulation) and six on type-3(combination of both). Participants were instructed to answer the questions after exploring the images. The questions were presented to participants in randomized order to eliminate the order bias for improving responses. Figure 2 shows image with visual manipulation technique.
**Procedure:** Participants were briefed about the experiment. For each participant, the eye tracker was calibrated using the Tobii calibration routine [15]. Participants were then asked to explore the images before answering the questions. The x-y coordinates of the gaze location and response to each question with timestamp were logged in a text file.
**Results:** For analysis, the images were divided into two regions. First region is the portion where deceptive patterns were present, and second region was without the deceptive element. Two dependent variables were considered – average time required to complete the task for each image, ratio of fixations in first region to total fixations in image. We undertook repeated measure ANOVA for dependent variables and found significant difference. We found significant main effect for total duration $F(11,110) = 1.252$, $p<0.05$, $\eta^2=0.111$ and, ratio of fixations $F(11,110) = 98.251$, $p<0.05$, $\eta^2=0.908$.

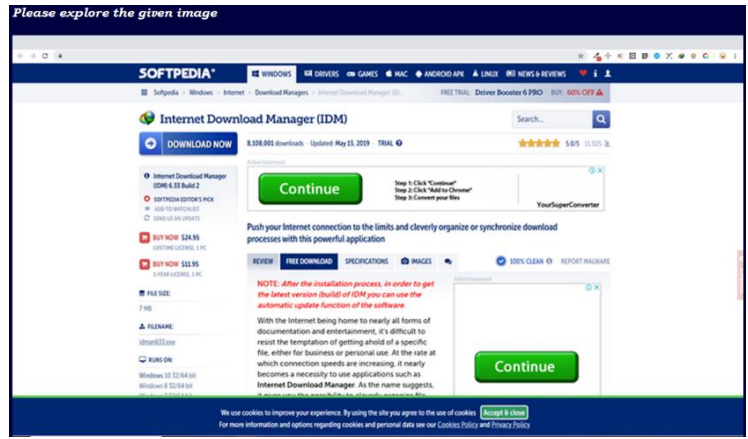

Fig 2: Example of website with visual manipulation

## 4. Applications

### 4.1 Digital twin sensor dashboard

A digital twin of Smart Factory layout inside Virtual Reality (VR) environment was modelled. Visualizations based on results of 2D graph study was used for analysing values of real-time sensors in VR environment. Visualization graphics was set up at locations of IoT nodes to embed real-time sensor readings on the virtual layout.

We integrated ambient light sensor (BH1750), temperature and humidity sensor (DHT22) to show real-time visualization of data streams in VR setup. Both sensors provide digital output. The BH1750 sensor has a built-in 16-bit A2D converter and output unit is lux. The DHT22 sensor provides temperature in celsius and humidity as relative percentage. Sensors were interfaced to the VR machine through their respective wireless modules [1]. After establishing a peer-to-peer connection, individual wireless module communicates with VR machine using UDP protocol at a frequency of 1 Hz.

The data stream taken from light sensor is visualized as an area graph. The graph shows change in light intensity over time and color of the graph changes if the value exceeds a threshold as shown in figure 3. Data obtained from temperature and humidity sensor is shown as separate circular bar. Similar to area graph, color of the circular bar also changes if value exceeds a threshold. Instantaneous values were converted to time-series values when user dwells using his/her eye gaze, providing a detailed view. Any abrupt change in sensor readings were reflected instantly via both visual and/or haptic feedback.

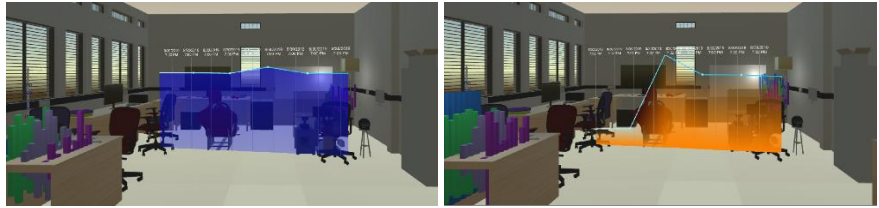

Fig 3: Area graph showing values from light sensors. The graph changes color after it exceeds a threshold value

## 5  Eye tracking data analysis tool

As discussed, cheaper eye tracker like Tobii Eye-X have restrictions in terms of research studies and high accuracy eye tracker is not always available. As a consequence, researcher use webcam eye tracker for studies in various domains [16,17]. In this paper we propose a web-based visualization tool to analyse single user eye tracking data. The tool is developed with the aim to provide an option for analysing eye gaze data to everyday user. A webcam-based eye tracking system is integrated with web-based interactive visualization techniques for analysing spatial-temporal aspects of the data. The tool is developed entirely on JavaScript. Webgazer.js library is used for the eye tracker and visualizations are developed with the help of D3.js library. The tool is hosted on python server and tested for Edge, Chrome, and Mozilla Firefox browser on Windows. User can undertake an eye tracking task and then immediately analyse the recorded data visually. Another option is to undertake a task and analyse the data later. Before starting eye tracking task, user has to undergo a calibration routine. The predefined calibration routine of Webgazer.js library is used in the tool. On completion of the task, user should be able to download the recorded data.

The downloaded data can then be analysed with web-based interactive visualization techniques. The tool allows K-means clustering on fixation position of the data. Fixation positions are shown in a scatter plot visualization. This technique would allow user to automatically identify relevant AOIs in the stimuli. K-Means clustering of eye gaze fixation is shown in figure 4. Heatmaps provide a quick glance on the data distribution over picture observed during an experiment or task. We developed an interactive heat map for this tool to analyse spatial attributes of the data. The kind of heatmap included in the tool is a fixation count. This means that each fixation of the user adds to the color on that position as shown in figure 5. Red color indicates high fixation count and green color indicates low fixation count. The tool also allows to change color gradients of the heatmap with text boxes in the control panel. Interactive gaze plot and scatter plot are also developed for analysing spatial-temporal aspects of the data. In general, scatter plots are used to show the spatial aspect of the data. This tool uses scatter plot with color-coding technique and interactivity to study the temporal aspect of the data. Gaze plots are similar to scatter plots but reveals more detailed knowledge about the data. Gaze plots can get cluttered and unorganized if number of fixations are large at a particular AOI. Interaction techniques are used with standard gaze plots to overcome this drawback. An interactive brush tool is used for displaying selected fixation points on both scatter plot and gaze plot.

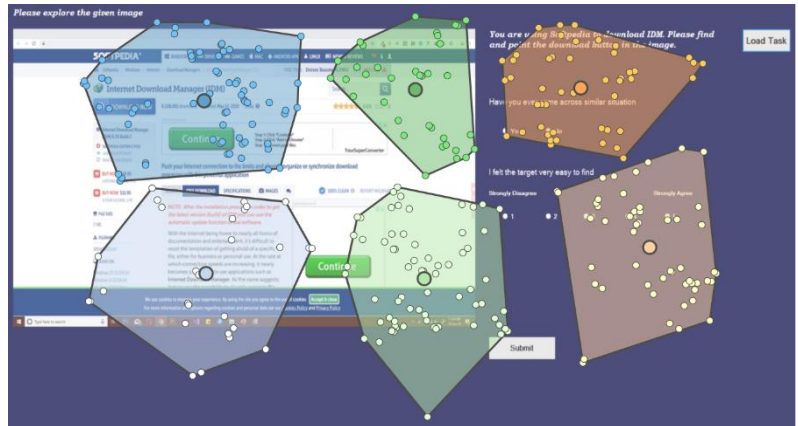

Fig 4: Clustering of fixation points on the stimulus

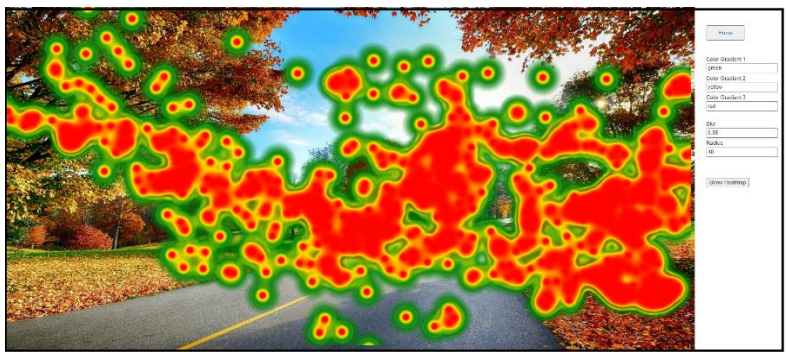

Fig 5: Heatmap of fixation count

## 5   Discussion

Our analysis on first case study found significant difference in gaze and response time for different graphs while performing similar tasks. Moreover, we analysed users' gaze fixation using expectation maximization algorithm. This analysis suggests that more fixations signifies more eye gaze movement, requiring longer duration to analyse data. We noticed that the average response time of bar graph is high and number of correct answers for area graph is low. This speed-accuracy tradeoff may lead us to future research questions. Our initial analysis consisted of dividing the screen in 9 regions and analysing first saccade positions and subsequent gaze movements from initial position. We noted users eye gaze first fixated on the central part and moving to bottom-center of the graph subsequently for all types of graph. Furthermore, we noticed that initial patterns for eye gaze fixations were similar for every graph however subsequent gaze movements were distinct for each graph. We found after analysing those movements that the sequence of bottom-center to middle-center and top-right to middle-center is most frequent among all two-region sequences. For analysis of our second study we investigated twelve web pages with three different types of deceptive patterns. A user study was undertaken on web pages that contained one of the three types of deceptive pattern and we found that fixations are independent of deceptive patterns used in the images. We also found that there is no significant correlation between fixations and task duration of the image.

## 6 Conclusion

This paper investigated visualization techniques and to discover suitable techniques for handling datasets. As demonstrated by results of the analysis, bar graph had the highest number of correct response and area graph had lowest response time. In our second study we investigated users' eye gaze behavior while interacting with deceptive materials in web pages. We analysed average task duration and ratio of fixations in deceptive region with entire region. We found that most of the images are significantly different from each other and fixations are independent of the deceptive patterns. We did not find any significant correlation between fixations and task duration of the image. We also developed a webcam-based visualization tool for analysing eye tracking data. The tool is independent of the implementation of webcam based eye gaze tracker and can generate similar visualization graph as reported in the study.